\documentclass{desyproc}

\usepackage[dvips]{graphicx}
\usepackage[euler]{textgreek}
\usepackage{subcaption}
\begin{document}
\title{ALPS\,II Status Report}

\author{{\slshape Aaron Spector$^1$ for the ALPS collaboration}\\[1ex]
$^1$DESY, Hamburg, Germany }

\contribID{Spector\_Aaron}

\confID{20012}  
\desyproc{DESY-PROC-2018-03}
\acronym{Patras 2018} 
\doi  

\maketitle

\begin{abstract}
ALPS II is a light shining through a wall style experiment that will use optical cavities to resonantly enhance the coupling between photons and axion-like particles in the mass range below 0.1\,meV. In the last year there has been significant experimental progress in the development of the optical system and the single photon detection schemes, as well as progress related to the preparation of the magnets and the on site infrastructure. 
\end{abstract}

\section{Introduction}

The Any Light Particle Search II (ALPS\,II)~\cite{alpstdr} is a light shining through a wall (LSW) style of experiment under construction at DESY in Hamburg, Germany. ALPS\,II will search for a broad class of low mass, weakly interacting, `axion-like' particles, that can mix with photons in presence of a strong magnetic field. The axion, the namesake for the axion-like particles, is a firm prediction \cite{Weinberg:1977ma,Wilczek:1977pj} of a solution to the strong CP problem in QCD proposed in 1977 by Peccei and Quinn \cite{Peccei}. 

LSW experiments are designed to measure the coupling between two photons and axion-like particles, $g_{a\gamma}$, in a laboratory setting~\cite{Redondo:2010dp}. This is done by shining a high power laser through a strong magnetic field thus generating a beam of axion-like particles that propagates through a wall that blocks the light. After the wall there is a second magnetic field that reconverts some of the axion-like particles to photons which are measured with a single photon detection scheme. Since LSW experiments generate the axion-like particles themselves, they can directly measure $g_{a\gamma}$. This is an advantage of this style of experiment over other axion-like particle searches such as haloscopes and helioscopes which rely on models of axions as dark matter or of their generation in the sun respectively.

ALPS\,II will take place in a section of the tunnel formerly occupied by the HERA experiment. Strings of ten 5.3\,T superconducting HERA dipole magnets will be used to generate the magnetic fields before and after the wall giving ALPS\,II 468 T$\cdot$m of magnetic field length in each of these regions. ALPS\,II will also be the first LSW experiment to use optical cavities before and after the wall to boost the probability that a photon will convert to an axion-like particle and then back to a photon after the wall. The power build up of the cavities, along with the long baseline, high magnetic field, and improvements in the detector technologies will help improve the sensitivity of ALPS\,II to the coupling between photons and axion-like particles over previous generations of LSW experiments by a factor of roughly 1000~\cite{Redondo:2010dp}. For masses below 0.11\,meV ALPS\,II will be able to detect axion-like particles down to couplings of $g_{a\gamma}\approx2\times10^{-11}\,\rm GeV^{-1}$~\cite{alpstdr} for a two week integration time.

\section{Optical system}

\begin{figure}[]
\centering
  \includegraphics[height=1.4in]{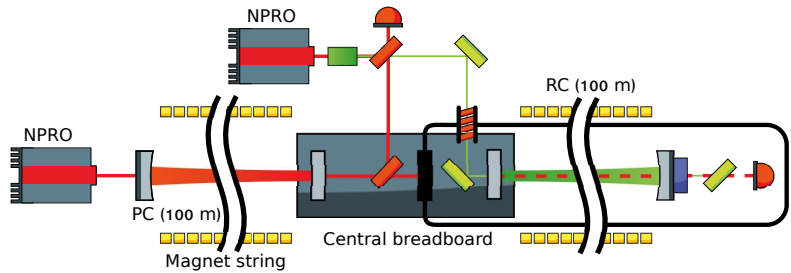}
\caption{ALPS\,IIc optical system.}\label{Fig:OS}
\end{figure}


The research and development for the ALPS\,II optics is taking place in two stages. The first, ALPS\,IIa, is a 20\,m testbed for the individual optical systems \cite{alpsIIa}. The second stage, ALPS\,IIc, represents the full scale 200\,m experiment and is shown in Figure~\ref{Fig:OS}. The setup is composed of two $\sim$100\,m optical cavities with the production cavity (PC) before the wall and the regeneration cavity (RC) after the wall. The PC will be seeded with a 30\,W fiber amplified nonplanar ring oscillator (NPRO) laser operating at 1064\,nm. With a power build up of 5000, the PC aims to achieve a circulating power of 150\,kW. The input optics will also be equipped with an automatic alignment system to maintain the coupling of the laser to the cavity. 50\,kW of power has already been demonstrated in the 9.2\,m ALPS\,IIa PC along with an automatic alignment system that allows the cavity to be stably operated for the anticipated measurement time of ALPS\,IIc.

The length of the RC must be resonant with the light circulating in the PC since the axion-like particles will have the same energy as the photons generating them. This requires stabilizing the differential length noise between the cavities to better than 0.5\,pico-meters. The sensing of the RC length will be performed with a 1064\,nm NPRO laser on the central optical table whose frequency is referenced to the light transmitted by the PC. This laser is then frequency doubled and the 532\,nm light is coupled into the RC. This is necessary since 1064\,nm light cannot be used as it would be indistinguishable from the regenerated photons. Because of this, the RC must be housed in a light-tight enclosure and requires dichroic mirrors that are reflective for both 1064\,nm and 532\,nm. The probability of axion-like particles reconverting to photon scales with the power build up of the RC for 1064\,nm. Therefore, this value is chosen to be as high as possible, at 40,000. The power build up for green light is significantly lower at 55 to avoid having a high circulating power.  This substantially lower power build up for green means that the light sensing the length of the RC must be stabilized to better than 1/10,000 of the cavity's linewidth.

In ALPS\,IIa the RC frequency control system has demonstrated the ability to maintain the resonance condition of a frequency doubled 1064\,nm NPRO laser to better than the required 0.5\,pm stability using the green light to probe the length of the cavity. Furthermore, a custom designed mirror mount composed of a piezoelectric actuator and a wave washer has demonstrated the ability to control the position of a 50\,mm mirror with a $\sim$4\,kHz control bandwidth and enough gain to suppress the environmental noise below the 0.5\,pm requirement. This demonstrates that it will be possible to operate ALPS\,IIc without any additional seismic isolation. A separate 1064\,nm was injected into the ALPS\,IIa RC and the power build up was measured to be 26,000$\pm$1,000.

The eigenmodes of the cavities must also share a spatial overlap of 95\% to ensure that the electromagnetic component of the axion-like particles couples to the RC. This requires that the eigenmodes of the cavities have less than a 5\,{\textmu}rad angular misalignment and less than a 1.3\,mm offset in their lateral position. To prevent any angular misalignment, the flat cavity mirrors at the center of the experiment will both be mounted to an in vacuum central optical bench (COB). The lateral position of the eigenmodes will be monitored with photodetectors mounted to the COB and this information will be fed back to three axis piezoelectric mounts that can actuate on the angle of the curved mirrors of the cavities. This will be used to maintain the lateral position of the eigenmodes with respect to the COB. Before a measurement run, the spatial overlap between the cavities can be checked by opening a shutter in the light-tight housing on the COB and allowing light transmitted by the PC to couple directly to the RC.

The alignment of the mirrors on the COB has been demonstrated by mounting mirrors to a test COB using Polaris\textsuperscript{\textregistered} mirror mounts. The alignment noise of the mirrors was measured using a Trioptics TriAngl\textcircled{e}r 300-57 auto-collimator. This measurement showed a peak-to-peak alignment drift of below 2\,{\textmu}rad over the course of a 5 day period.

\section{Detection Systems}

ALPS\,II will have the benefit of using two independent single photon detection systems to confirm the results of the experiment. One of these systems, known as the transition edge sensor (TES), is a microcalorimeter capable of measuring the temperature change induced by incident photons. The other measurement system uses a heterodyne detection technique to measure the optical interference signal between the regenerated photons and a local oscillator.

The TES detection system exploits a thin tungsten film stabilized at its critical temperature between normal and superconductivity by a bias current. It is read out via a two stage SQUID amplifier~\cite{Jan}. When photons are absorbed by the film they will increase its temperature slightly for a brief period of time. This will in turn cause a change in the resistance of the tungsten and lead to a sudden drop in current. At the moment the cryostat has been successfully set up and the rest of the system is being optimized.

The heterodyne detection technique measures the interference beatnote between the regenerated photons and a local oscillator. By exactly knowing the relative frequencies of these fields, the measured data can be demodulated at the signal frequency. 
This concept relies on the phase coherence between the regenerated field and the local oscillator which requires that the local oscillator is phase coherent with the light circulating in the PC. A demonstration of this system showed that it was capable of measuring a signal with a power of 1 photon per 30\,s with no measureable background over the course of two weeks~\cite{Bush}. For a more detailed overview and status report please see the article in these proceedings by G. Messineo.

\section{Site preparation and timeline}

\begin{figure}[t]
\centering
\begin{subfigure}{.5\textwidth}
  \centering
  \includegraphics[height=1.7in]{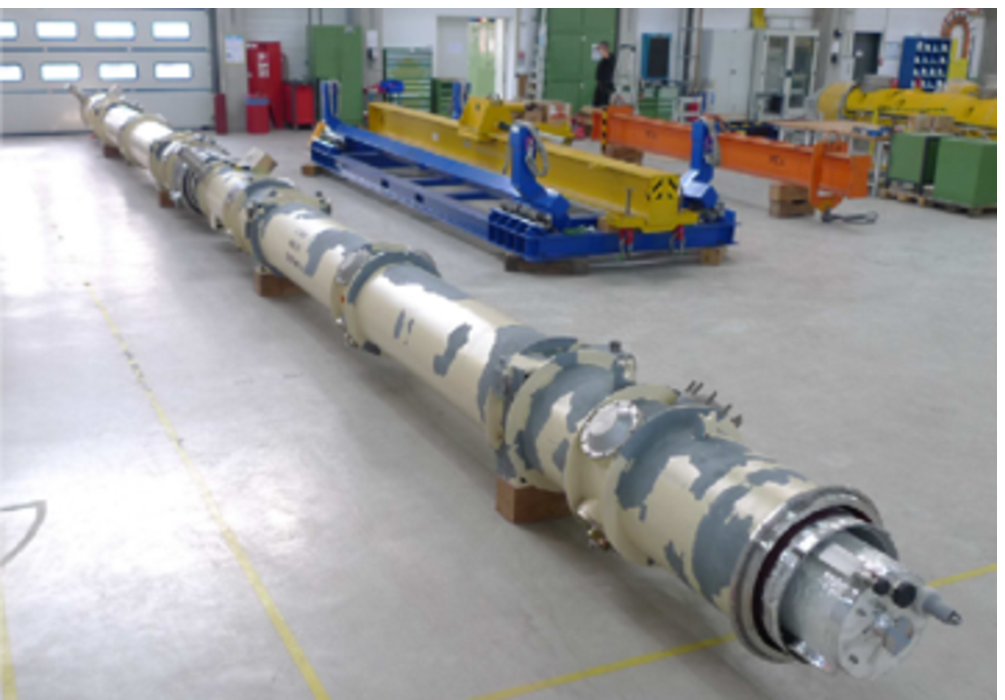}
  \caption{}
  \label{fig:sub1}
\end{subfigure}%
\begin{subfigure}{.42\textwidth}
  \centering
  \includegraphics[height=1.7in]{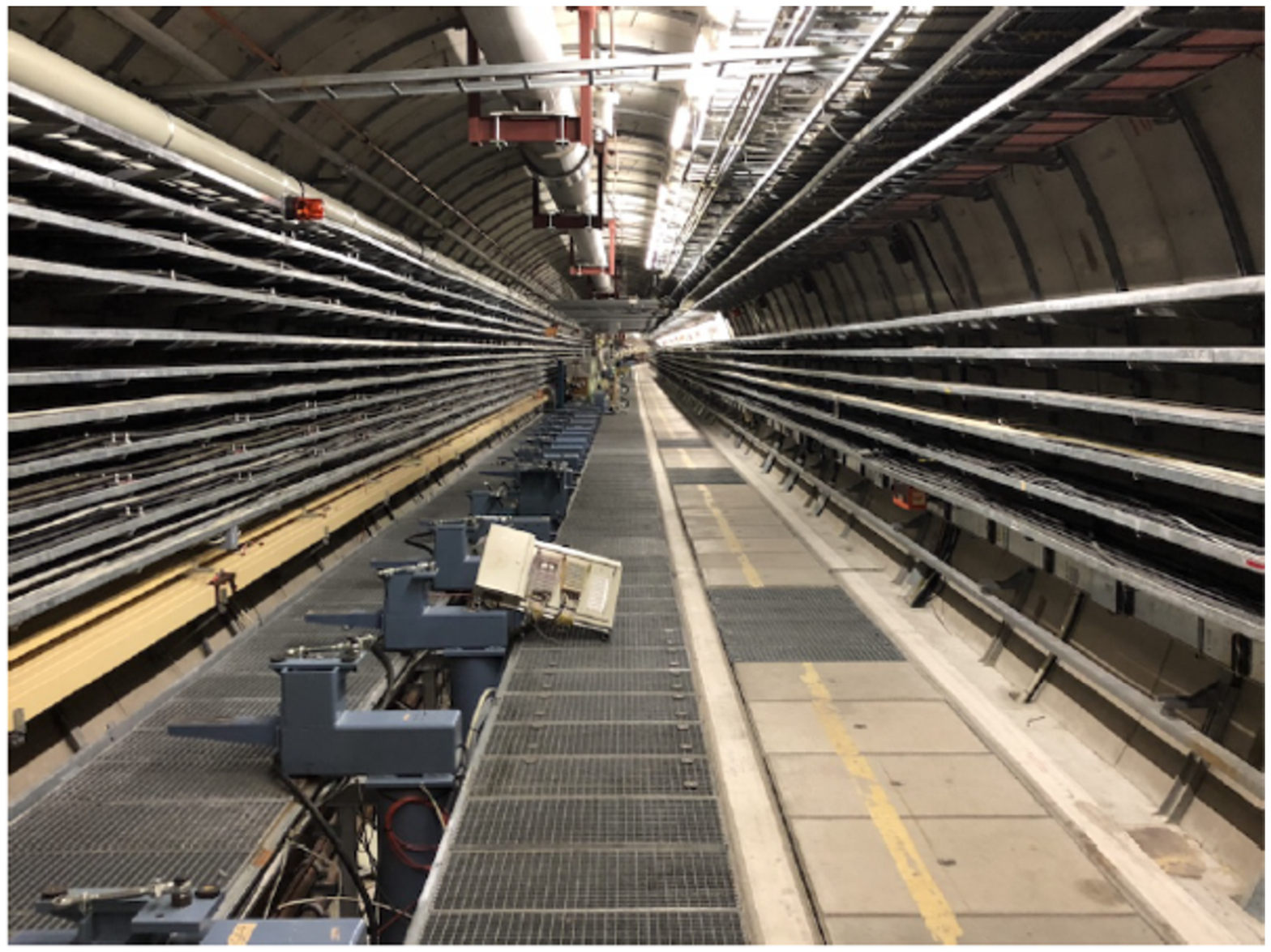}
  \caption{}
  \label{fig:sub2}
\end{subfigure}
\caption{(a) Test string of two magnets. (b) The cleared HERA tunnel. }
\label{fig:test}
\end{figure}

As mentioned earlier, ALPS\,IIc will require twenty superconducting HERA dipole magnets. Since these magnets were originally used in the arcs of the HERA accelerator, they must be unbent to provide sufficient aperture for the cavity eigenmodes. At the moment, sixteen of the magnets have been unbent and successfully operated. Furthermore, as Figure~\ref{fig:sub1} shows, two magnets were connected in a test string to confirm that the interconnections between the straightend magnets function properly. The tunnels and hall that will house ALPS\,IIc have also been cleared in preparation for the experiment as seen in Figure~\ref{fig:sub2}.


The construction of the cleanroom for the first laser area of ALPS\,IIc is scheduled for fall 2019. The installation of the first string of ten magnets for the PC should be completed by early 2020 with the installation of the second string of ten magnets for the RC finished by summer 2020. 
The installation of the full scale optical system will be completed during the fall of 2020, and the ALPS collaboration is looking forward to begin taking data at the end of 2020.

\section*{Acknowledgments}

The ALPS Collaboration would like to thank the DESY accelerator division, especially the MEA, MKS and MVS groups for their crucial support. This work is financially supported by the Helmholtz Foundation, the Deutsche Forschungsgemeinschaft, the Heising-Simons Foundation under Grant No. 2015-154, and the NSF under Grants No. 1505743 and No. 1802006.


\begin{footnotesize}
\begin{footnotesize}

\end{footnotesize}

\end{footnotesize}



\begin{thebibliography}{99}
%
\bibitem{alpstdr}
R.~B{\"a}hre, {\it et al.},
  ``Any light particle search {II} --- {T}echnical {D}esign {R}eport,"
  J. Instrum. \textbf{8}, T09001 (2013) doi:10.1088/1748-0221/8/09/T09001
    [arXiv:1302.5647 [physics.ins-det]].
	

\bibitem{Weinberg:1977ma}
  S.~Weinberg,
  ``A New Light Boson?,''
  Phys.\ Rev.\ Lett.\  {\bf 40} (1978) 223.
  doi:10.1103/PhysRevLett.40.223

\bibitem{Wilczek:1977pj}
  F.~Wilczek,
  ``Problem of Strong  $P$  and  $T$  Invariance in the Presence of Instantons,''
  Phys.\ Rev.\ Lett.\  {\bf 40} (1978) 279.
  doi:10.1103/PhysRevLett.40.279

\bibitem{Peccei} 
   R.~Peccei, and H.~Quinn.,
  ``CP Conservation in the Presence of Instantons,''
  Phys.\ Rev.\ Lett. {\bf 38}, 1440 -- 1443 (1977)
  doi:10.1103/PhysRevLett.38.1440
	
\bibitem{Redondo:2010dp}
  J.~Redondo and A.~Ringwald,
  ``Light shining through walls,''
  Contemp.\ Phys.\  {\bf 52} (2011) 211
  doi:10.1080/00107514.2011.563516
  [arXiv:1011.3741 [hep-ph]].
	
	
\bibitem{alpsIIa}
A.~Spector, {\it et al.}, ``Characterization of optical systems for the ALPS II experiment,"
 Opt.\ Express \textbf{24} 9237 (2016) doi:10.1364/OE.24.029237
 [arXiv:1609.08985 [physics.optics]].
 
 \bibitem{Jan}
  J.~Dreyling-Eschweiler {\it et al.},
  ``Characterization, 1064 nm photon signals and background events of a tungsten TES detector for the ALPS experiment,''
  J.\ Mod.\ Opt. {\bf 62}, 1132 -- 1140 (2015) doi:10.1080/09500340.2015.1021723
  [arXiv:1502.07878 [physics.ins-det]].

 
 \bibitem{Bush}
Z.~Bush, {\it et al.}, ``Coherent detection of ultraweak electromagnetic fields,"
 Phys.\ Rev.\ D \textbf{99} 022001 (2019) doi:10.1103/PhysRevD.99.022001
 [arXiv:1710.04209 [physics.ins-det]].
	

\end{thebibliography}
\end{document}